\documentclass[11pt,twoside,dvips]{article}
\usepackage{graphicx}
\usepackage{verbatim}
\usepackage{amssymb}
\usepackage{amsfonts}
\usepackage{amsmath}
\usepackage{pifont}
\usepackage[spanish,english]{babel}
\usepackage{verbatim}
\usepackage{epsfig}
\usepackage{bm}
\usepackage{longtable}
\usepackage{fancyhdr}
\usepackage{hyperref}

\begin{document}

\fancypagestyle{plain}{%
\fancyhf{}%
\fancyhead[LO, RE]{XXXVIII International Symposium on Physics in Collision, \\ Bogot\'a, Colombia, 11-15 September 2018}}

\fancyhead{}%
\fancyhead[LO, RE]{XXXVIII International Symposium on Physics in Collision, \\ Bogot\'a, Colombia, 11-15 september 2018}

\title{Ultrahigh-energy cosmic rays}
\author{ Analisa G. Mariazzi $\thanks{%
e-mail: mariazzi@fisica.unlp.edu.ar}$
\\ Instituto de F\'isica La Plata, CONICET, 
\\ for the Auger Collaboration$\thanks{%
e-mail: auger\_spokespersons@fnal.gov}$
\\ Pierre Auger Observatory, Malarg\"ue, Argentina \\ (Full author list: \url{http://www.auger.org/archive/authors_2018_09.html})
        }
\date{}
\maketitle

\begin{abstract}
Ultrahigh-energy cosmic rays (UHECRs) arrive at Earth from the most energetic astrophysical accelerators in the universe. 
They collide with atoms in the upper atmosphere with energies about ten times higher than any man-made accelerator, and produce gigantic cascades of secondary particles, called extensive air showers (EAS). 
Extensive air showers can be detected spreading particle detectors over a large area to record the interactions of secondary particles.

The Pierre Auger Observatory has been designed to investigate the origin and nature of UHECRs using the combination of information from a surface array, measuring the lateral distributions of secondary particles at the ground, and fluorescence telescopes, observing the longitudinal profile of the electromagnetic component of EAS, providing an enhanced reconstruction capability. 

In this contribution, the status and prospects of understanding the physics of UHECRs will be reviewed, focusing on the progress made thanks to the measurements of the Pierre Auger Observatory.
Physics results from the ultrahigh-energy cosmic ray data collected with the Pierre Auger Observatory opened new perspectives and motivated an upgrade of the Observatory, AugerPrime, whose main characteristics are also presented.
\end{abstract}
\section{Introduction}
Until the advent of high-energy particle accelerators, cosmic rays provided a way to discover new particles, like the positron, the muon, the pion, the kaon and several more. 
Accelerators provide now the best hunting ground for new particles, but the physics of cosmic rays is still widely studied.

The energies of the primary cosmic rays range from GeV energies to as much as $10^{20}$ eV, far higher than the maximum energies attainable at current man-made particle accelerators like the LHC. 
The flux of the particles spans from around one particle per square meter per second at GeV energies to less than one particle per square kilometer per century at the highest energies around $10^{20}$ eV.

In the past, two kind of mechanisms for the origin of UHECRs were proposed: the so-called top-down mechanism like, for example, the decay of very massive long–lived particles being produced in the early universe, and the so called bottom-up mechanism, where particles are accelerated in astrophysical sources like acceleration in front shocks. 
The maximum energy attained in a conventional accelerator is given by the atomic number $Z$ times the size of the accelerating object times the magnetic field $B$. 

As ultrahigh-energy cosmic rays propagate from their sources to Earth, they interact with the intergalactic medium, suffering energy losses that affect the initial energy spectrum and mass composition.
The most important of these processes involves the interaction with the low-energy photons of the cosmic microwave background.
For protons, the pion photo-production occurs above a threshold energy of ∼60 EeV, called the GZK cutoff. Cosmic-ray protons should then not exceed this energy limit. The photo-disintegration of heavy cosmic-ray nuclei would have a similar effect.
The observed cosmic rays above the GZK cutoff should then come from some nearby sources in the local universe, which have yet to be identified. An anisotropy in the arrival direction distribution of cosmic rays in the energy range of the GZK suppression is expected, due to the small propagation distance 
and the highly anisotropic matter distribution in the nearby universe. 

Above $10^{14}$ eV, direct measurements with balloon or satellite experiments are not possible and the properties of the primary cosmic ray have to be inferred indirectly from the air shower observables measured using particle detectors at the ground or the air-fluorescence technique. 

The Pierre Auger Observatory is the world’s largest air shower experiment. It has a unique and powerful design based on the simultaneous detection of air showers by combining independent and complementary instruments: the fluorescence detector (FD) and the surface detector array (SD). 

While the SD is sensitive to the electromagnetic and muonic shower component and its duty cycle is almost 100\%, the FD measures the calorimetric energy deposited in the atmosphere and its operation is only possible in clear moonless nights with a duty cycle of about 13\%.

The SD is composed of 1660 water-Cherenkov detectors on a triangular grid of 1500 m spacing, which covers an area of 3000 km$^2$. 
The array is fully efficient for energies above 3.$10^{18}$ eV.

There is an infill array in a small area of the array with a denser spacing (750 m) that increases the SD sensitivity towards lower energies.
The infill array is fully efficient above 3.$10^{17}$ eV. 

The FD consists of 24 air fluorescence telescopes  grouped in four sites, that overlook the area covered by the SD.
Three additional telescopes pointing at higher elevations (HEAT) are located in the infill region near one of the FD sites to detect lower energy showers that develop higher in atmosphere.

The advantage of a hybrid detector relies on the fact that it is possible to calibrate the SD signal using the events simultaneously recorded by the FD (hybrid events) to reconstruct the energy of SD-only events,  
avoiding to a large extent the use of model-dependent Monte Carlo simulations and gaining better control of the systematic uncertainties in the energy scale.
\section{Search for neutral messengers}
\begin{figure*}
\includegraphics[width=84mm]{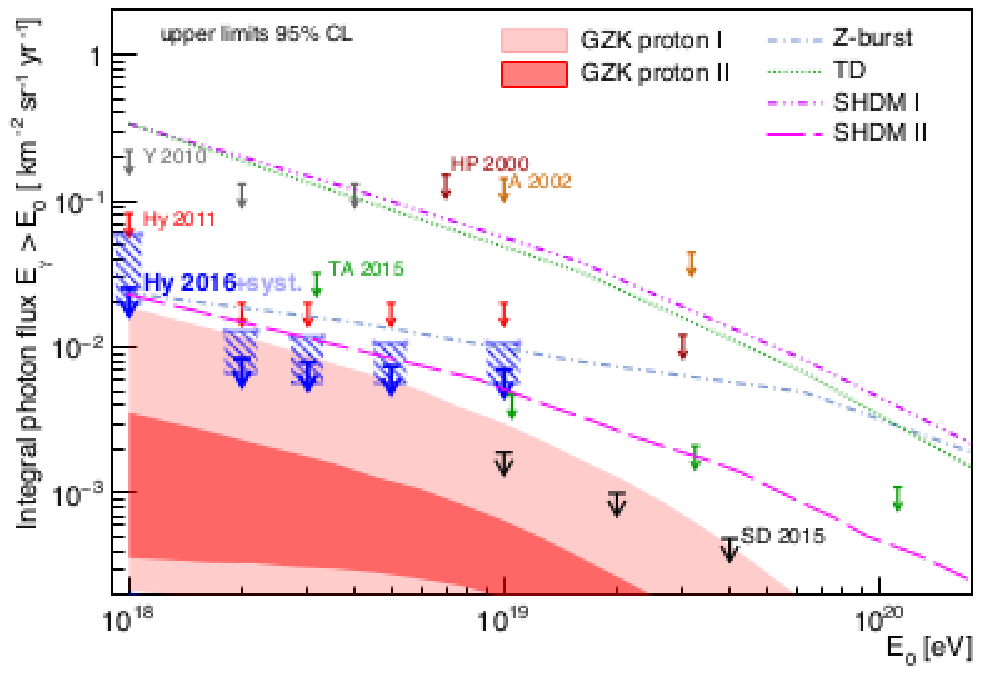}
\includegraphics[width=84mm]{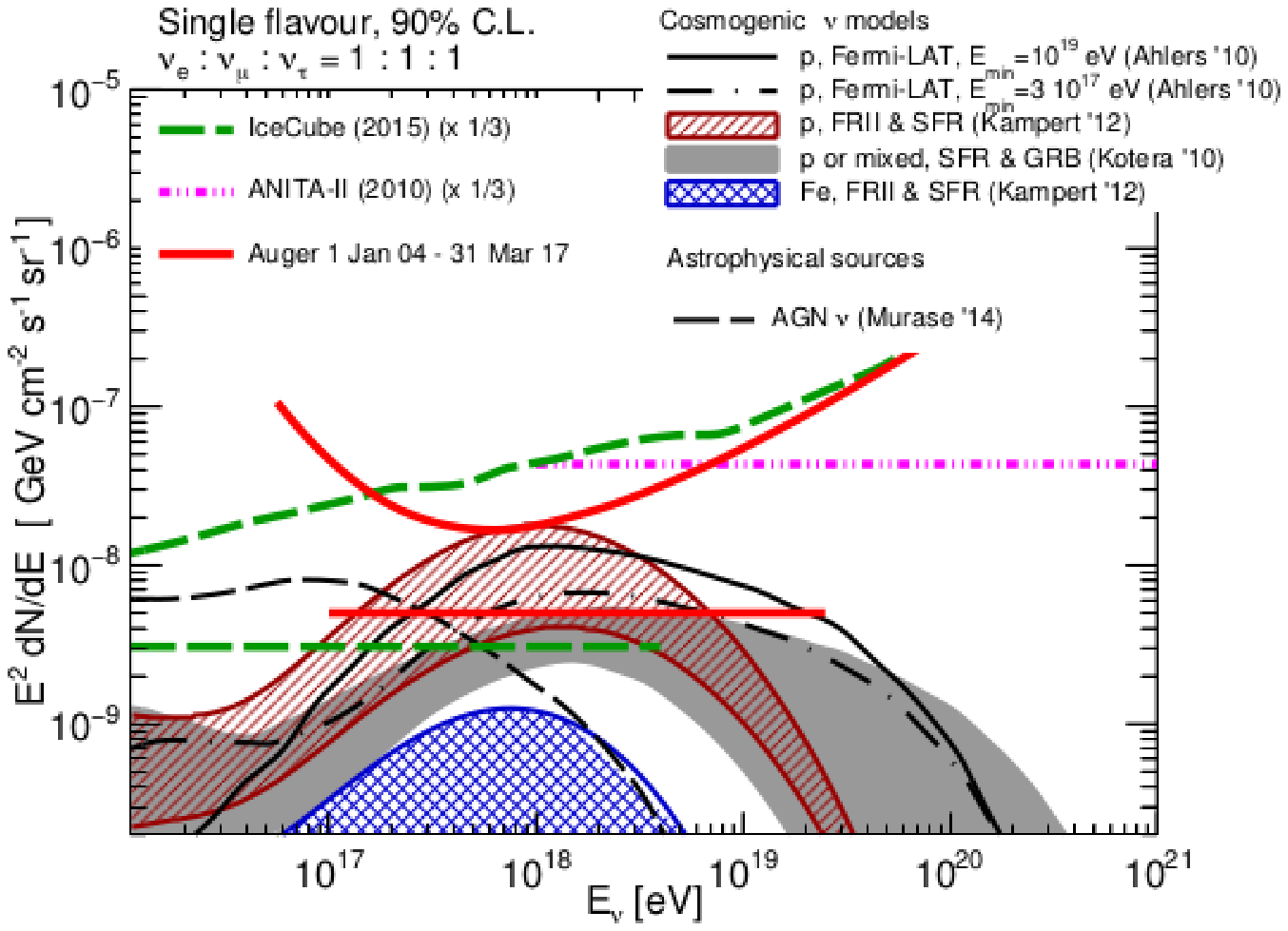}
\caption{Upper limits on the diffuse photon flux (left) and neutrino flux (right) compared to limits from other experiments and model predictions~\cite{photon,neutrino}.}
\label{fig:neutral}
\end{figure*}
Neutral messengers, like photons and neutrinos, are the best probe to point an UHECRs production source, as they are not deflected by galactic or extragalactic magnetic fields. 

Many current models for the origin of UHECRs  predict a flux of photons and neutrinos created as secondary particles when the cosmic rays attain their ultra-high energies or when they propagate through the Universe.
The diffuse flux of these neutral messengers then carries information about the propagation of cosmic rays, their mass and the spatial distribution of the source. 

The identification of photons relies on the fact that photon-induced showers develop deeper in the atmosphere, produce a lower average number of muons and, having a steeper lateral distribution function, are characterized by a smaller footprint on ground. 

Neutrino-induced air showers are identified by searching for upward-going near horizontal events (Earth-skimming neutrinos) or down-going near horizontal events with a larger electromagnetic component (first interaction occurs very deep in the atmosphere)\cite{neutrino}. 

Results on the diffuse flux of high energy photons and neutrinos are shown in Figure~\ref{fig:neutral}.

The absence of neutrino candidate events yields bounds on the diffuse flux of UHE neutrinos and constraints on models attempting to explain the origin of UHECRs (see Figure \ref{fig:neutral}-(right)).
Some cosmogenic models are already excluded (at 90 \% C.L.), particularly those that assume primary protons and a strong source evolution with redshift.

The Auger Observatory has set photon limits with both the hybrid and SD detection methods, seen Figure \ref{fig:neutral}-(left). 

A discovery of a substantial photon flux could have been interpreted as a signature of top-down models. Instead, strong flux limits have been placed, disfavoring top-down models in which is assumed that UHECRs are the decay products of Super Heavy Dark Matter (SHDM), topological defects (TD) or $Z0$ bosons created in the interaction of extremely high-energy neutrinos with the relic neutrino background (Z-burst).
\section{Energy spectrum}
\begin{figure*}
\includegraphics[width=70mm]{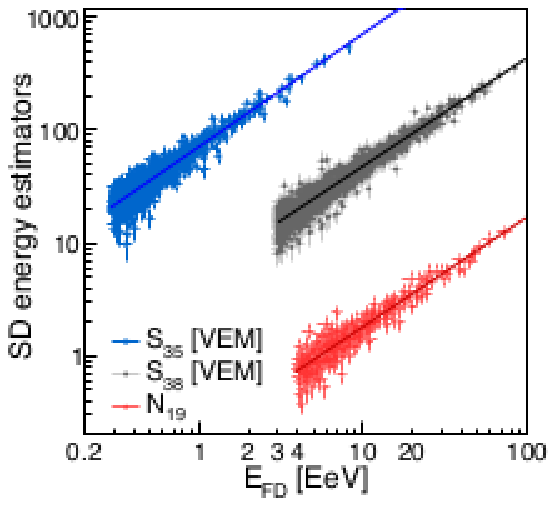}
\includegraphics[width=95mm]{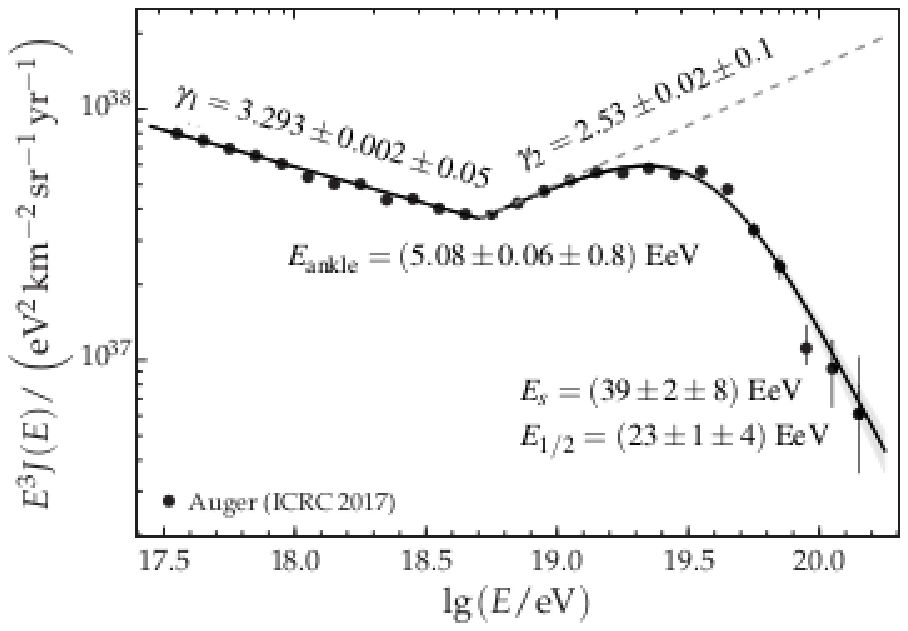}
\caption{Left: Energy calibration of the SD. The shower size measured for vertical events with the SD-1500 ($S_{38}$) and SD-750 ($S_{35}$) array and for inclined showers ($N_{19}$) is shown as a function of the energy measured with the FD ($E_{FD}$).
Right: Combined energy spectrum obtained with the four spectrum components (FD hybrid, inclined SD, vertical SD-1500 and SD-750). 
The line shows a fit to the spectrum with a broken power law and a suppression at ultrahigh energies. The gray dashed line indicates the same broken power law without suppression. The fitted spectral indices and  energies of the break and suppression are superimposed together with their statistical and systematic uncertainties\cite{spectrum}.}
\label{fig:spectra}
\end{figure*}
The shape of the spectrum is linked to the origin and nature of UHECRs. 
Due to the hybrid nature of the Auger Observatory one can determine the energy spectrum of primary cosmic rays without strong dependence on the model assumptions about hadronic interaction models in air showers. 

From a selected sample of hybrid events, a calibration curve is obtained, which is then used to find energies of SD events.

The observable chosen to characterize the size of an SD event for the SD-1500 array is the signal at 1000 m from the shower axis, normalized to median zenith angle of the events of 38$^\circ$ (S38); for the SD-750 array is the signal normalized to a median zenith angle of the events of 35$^\circ$ (S35), while for inclined showers (with the zenith angle larger than 60$^\circ$ seen by SD-1500 array) the $N_{19}$ energy estimator is used \cite{n19}.
The corresponding calibration curves are shown in Figure \ref{fig:spectra}-(left).

The energy spectrum of UHECRs is measured at the Pierre Auger Observatory in four distinct ways. 
The four datasets include data from: FD Hybrid, SD 750 m array and SD 1500 m array 
in total comprise more than 300,000 events recorded over the course of more than ten years, with a sky coverage ranging from $-90^\circ$ to $+45^\circ$ in declination.

The individual spectra obtained from the four datasets agree within their uncertainties, hence a combined spectrum can be determined, as it is shown in Figure \ref{fig:spectra}-(right). 
The dominant systematic uncertainty on the combined spectrum stems from the overall uncertainty on the energy scale of 14\%. 
With the combined spectrum, the spectral features, i.e. the ankle and the flux suppression at the highest energies, can be determined with unprecedented precision\cite{spectrum}. 

The flux suppression above 40 EeV has been observed with more than 20$\sigma$ significance, but its origin is still unclear. It could be due to a propagation effect (GZK) or by the exhausted power of the cosmic accelerators or a combination of both. 
The origin of the suppression cannot be determined without the knowledge of mass composition in this region, which is a keystone to clarify these enigma.
\section{Mass composition}
\begin{figure*}
\includegraphics[width=150mm]{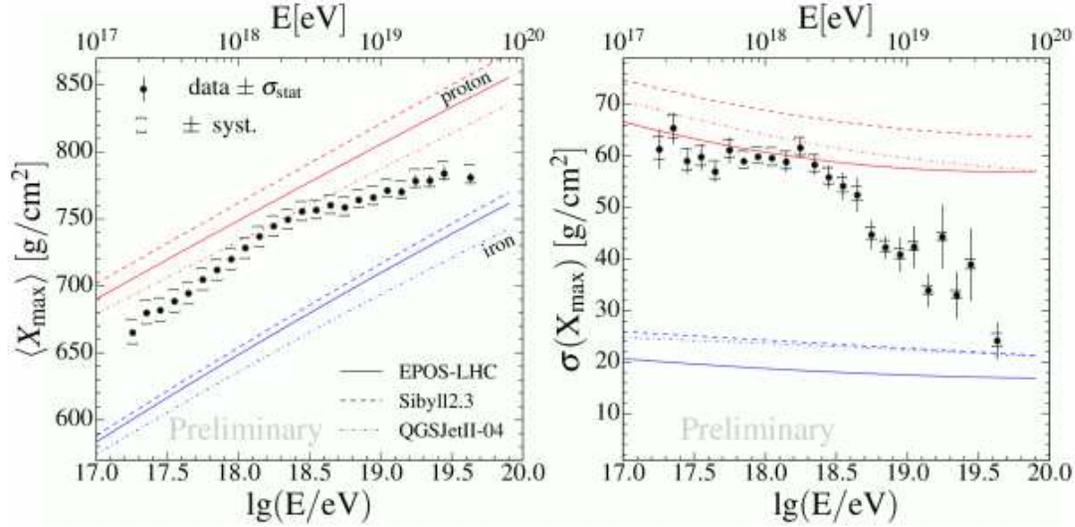}
\caption{The mean (left) and the standard deviation (right) of the measured $X_\text{max}$ distributions as a function of energy compared to air-shower simulations for proton and iron primaries\cite{mass}.}
\label{fig:mass1}
\end{figure*}
The most direct observable to infer the mass composition of cosmic rays from air-shower measurements is the atmospheric depth of the shower maximum $X_\text{max}$ measured using the fluorescence detector (FD). 

The average depth of shower maximum changes with energy in an unexpected way in the data from the Pierre Auger Observatory.
Around 3.$10^{18}$ eV, it shows a distinct change of $\langle X_\text{max} \rangle$ with energy and the shower-to-shower variance 
$\sigma(X_\text{max})$ decreases \cite{mass}. 
This trend implies a gradual shift to a heavier composition when interpreted with the leading LHC-tuned air-shower models. 
\begin{figure*}
\includegraphics[width=130mm]{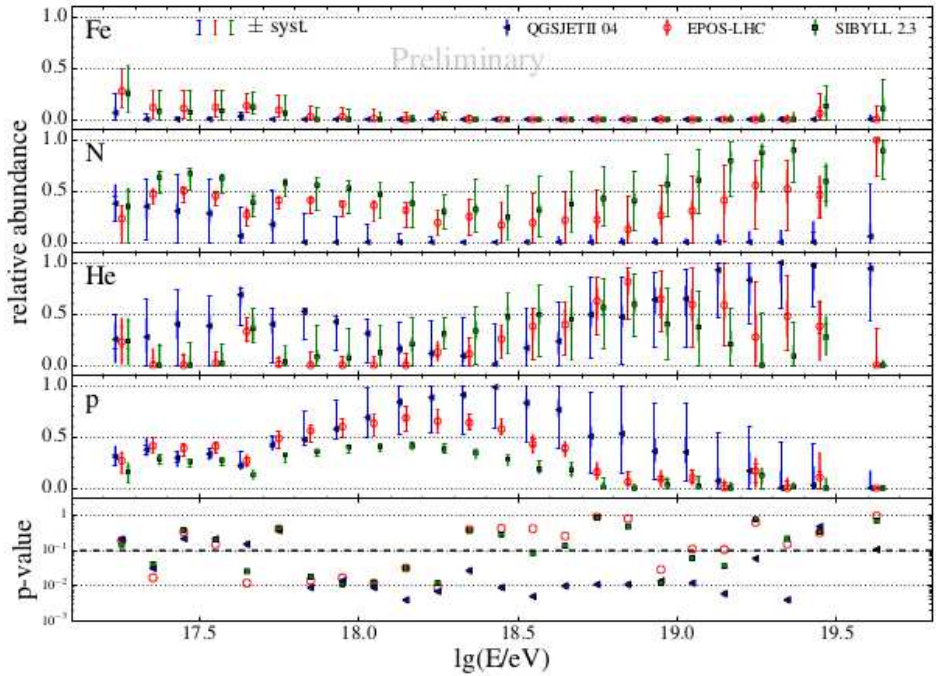}
\caption{Mass-fraction fits obtained using parameterizations of the  $X_\text{max}$ distributions from the fluorescence  $X_\text{max}$ data.
The error bars indicate the statistics (smaller cap) and the systematic uncertainties (larger cap). The bottom panel indicates the goodness of the fits (p-values)~\cite{mass}.}
\label{fig:mass2}
\end{figure*}

As an additional analysis approach, the shape of the $X_\text{max}$ distribution was further analyzed \cite{mass}: the data were fitted to various mass fractions for several hadronic interaction models. 
The best fit was obtained with four components included, p, He, N, Fe, and the relative proportions appear to be evolving with energy, as shown in Figure \ref{fig:mass2} for three different high-energy hadronic interaction models. 
Protons are most abundant near the ankle region, while helium dominates above $10^{19}$ eV and the fraction of iron nuclei is insignificant in the whole energy range. However, it should be pointed out that the interpretation of the mass fractions is model dependent.

Due to the limited duty cycle of the FD, the data do not cover the flux suppression region, where there are very few events making difficult mass composition studies with this detector.
\section{Arrival directions of cosmic rays}
\begin{figure*}
\includegraphics[width=100mm]{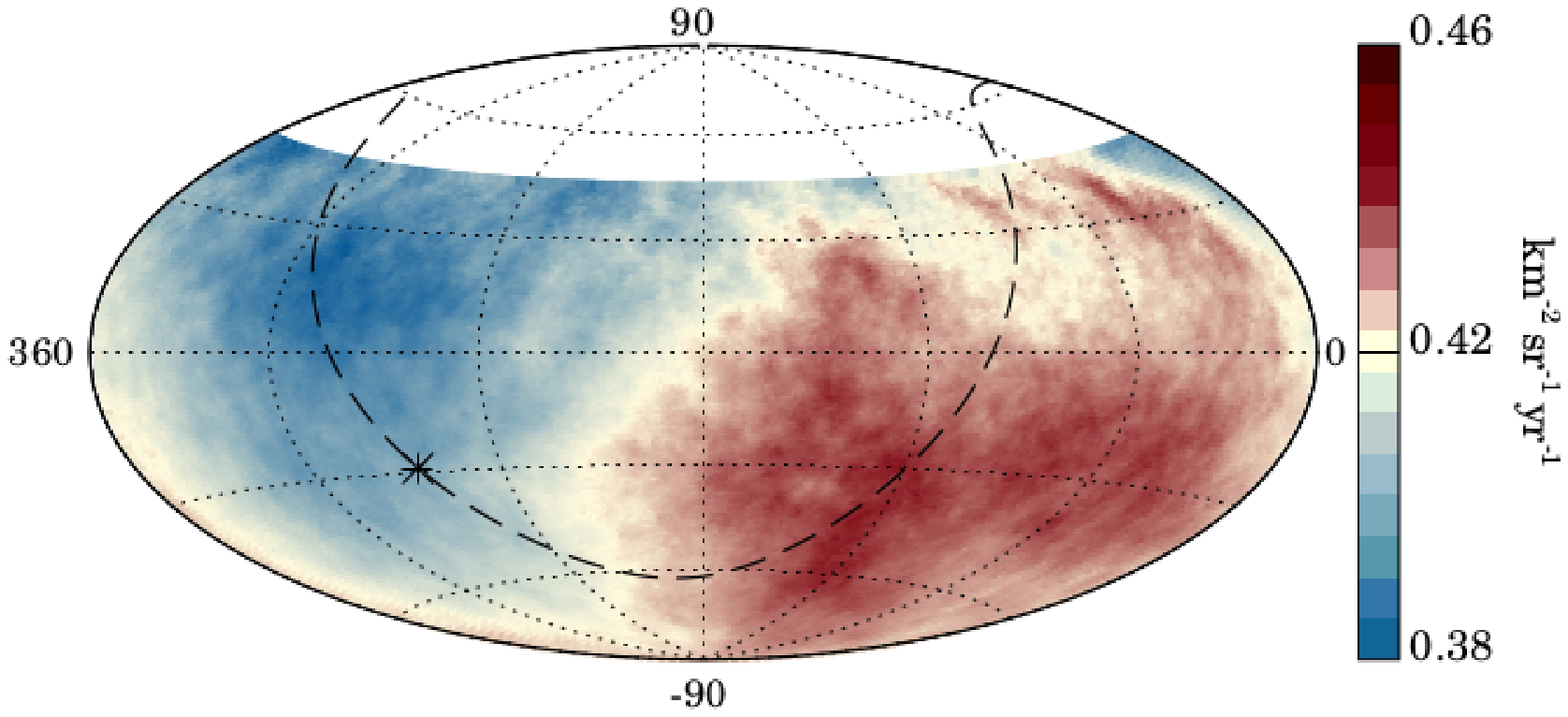}
\includegraphics[width=60mm]{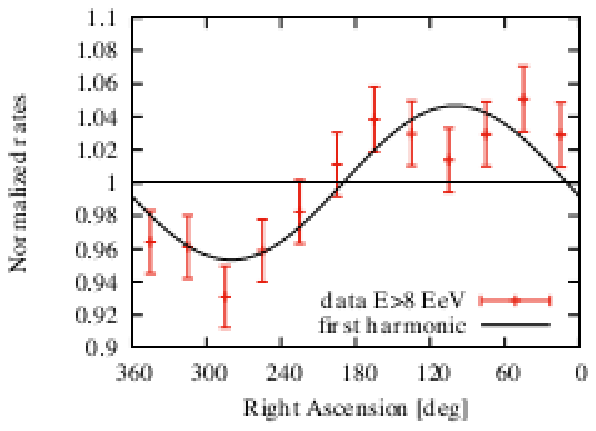}
\caption{ 
Right: Distribution of the normalized rate of events with E $\ge$ 8 EeV as a function of the right ascension\cite{anis}. The function corresponding to the first harmonic is also indicated with a solid line, showing that the distribution is compatible with a dipolar modulation. 
Left: Sky map in equatorial coordinates, Hammer projection, showing the UHECR flux above 8 EeV smeared by a 45$^\circ$ top-hat function. The dashed line represents the Galactic plane and the asterisk marks the Galactic center\cite{anis}.
}
\label{fig:anis}
\end{figure*}
The study of the distribution of the arrival directions can reveal the sources of UHECRs.

Since cosmic rays are charged particles, they are deflected 
when passing through the Galactic and extragalactic magnetic fields.
One can either look at the highest energies, where deflection should be the least and search for small and intermediate scale anisotropies or look also at lower energies looking for large-scale anisotropies. 

Indications of an anisotropic distributions of sources were found both at large angular scale and at intermediate angular scale.

The distribution at large angular scales of the arrival directions of cosmic rays is studied performing Rayleigh analysis of the first harmonic in right ascension.
Data obtained since the start of data taking with the Observatory
has been divided into two energy bins (4 EeV $< E \leq$ 8 EeV and $E \geq$ 8 EeV) and they were analyzed to determine the amplitude of the first harmonic in right ascension.

In the lower energy bin (4-8 EeV) the results are consistent with an isotropic distribution, but for $E \geq$ 8 EeV a significant modulation is observed. In this higher energy range, the amplitude of the first harmonic was found to be incompatible with expectations from an isotropic distribution at more than 5.2 $\sigma$ level~\cite{anis}. 

The distribution of events with $E \geq$ 8 EeV, in equatorial coordinates, is shown in Figure \ref{fig:anis}-(left).
The distribution of the normalized rate of events with $E \geq$ 8 EeV as a function of the right ascension is shown in Figure \ref{fig:anis}-(right). The solid line indicates the function corresponding to the first harmonic, showing that the distribution is compatible with a dipolar modulation. 

The origin of the discovered dipole remains a subject of speculation at this point.  
It can however be concluded that, as the direction of the reconstructed dipole lies about 125$^\circ$ away from the Galactic Centre, this provides strong support to the hypothesis of an extragalactic origin of the highest-energy cosmic rays.

In a recent study~\cite{intanis}, a comparison of the measured arrival-direction distribution with the distribution of objects from selected catalogs was done. 
Two different candidate sources were taken into account: AGNs ($\gamma$AGN) and starburst
galaxies (SBG). For these two catalogs, a flux map was constructed which was then used in a likelihood ratio analysis to test the deviation from isotropy in data. 
While the significance for the AGNs is close to the one obtained in a previous analysis (2.7$\sigma$), a much higher significance is obtained with the newly tested starburst hypothesis. For the starburst galaxies, isotropy is disfavored at the 4$\sigma$ level, which indicates an excess of events from strong nearby sources.
\section{Tests of Hadronic Interactions at Ultrahigh Energies}
The hybrid nature of the Pierre Auger Observatory offers a great capability for reconstructing a number of properties of EAS that can be used to test the consistency of the EAS description with the current hadronic interaction models at energies beyond that accessible in the laboratory.

\begin{figure*}
\includegraphics[width=130mm]{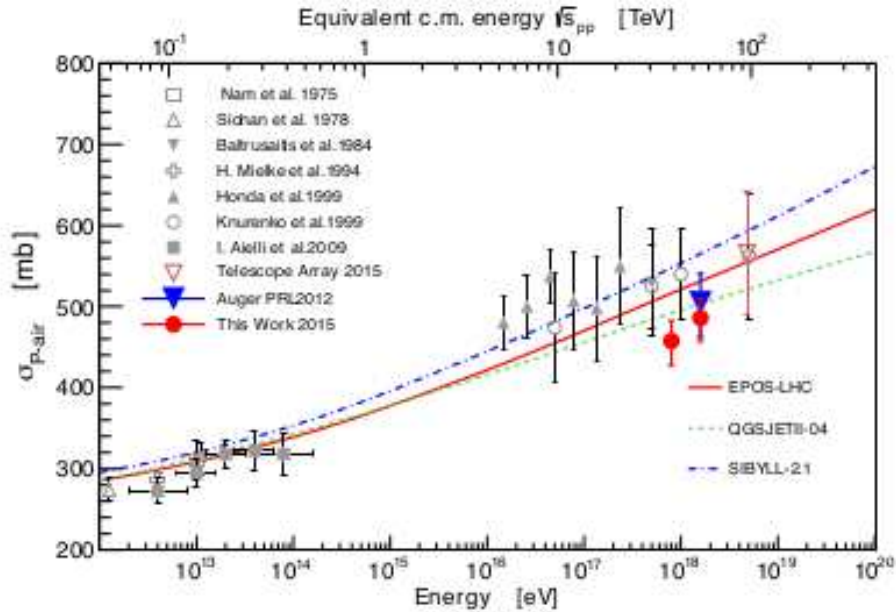}
\caption{Measurement of p-air cross-section compared to previous data and model predictions \cite{cross}.}
\label{fig:cross}
\end{figure*}

The tail of the $X_\text{max}$ distribution is sensitive to the proton-air cross-section.
A data sample enriched in protons could be obtained by analysing only the most deeply penetrating events. 
The proton-air cross-section has been derived analysing the tail of the distribution of $X_\text{max}$ observed with the Pierre Auger Observatory in two energy bins around $10^{18}$ eV \cite{cross}. 
Results presented in Figure \ref{fig:cross} are consistent with a rising cross section with energy. However, the statistical precision is not yet sufficient to make a statement on the functional form of the cross-section extrapolations that is used in various models of hadronic interactions.

Measurements of the muon content of extensive air showers at the Pierre Auger Observatory could give some insights into the physics of hadronic interactions and multiparticle productions beyond LHC energies. 
The hadronic cascade is the main engine that drives the development of nuclei-induced EAS.
Muons are mostly produced by the decay of charged mesons (that were feeding the hadronic cascade in every generation until they decay into muons), becoming direct messengers from the hadronic skeleton of the shower.

An efficient way of measuring muons is using highly inclined showers, since the electromagnetic component of the air shower is almost completely absorbed in the atmosphere while most muons still penetrate down to the ground.

Hybrid events with zenith angle $62^\circ < \theta < 80^\circ$ were selected. 
The reconstruction is based on the fact that the muon number distribution at the ground can be described by a density-scaling factor that depends on the shower energy and primary mass, and by  a lateral shape that, for a given  arrival direction $(\theta,\phi)$ of the shower,  is consistently reproduced by different hadronic interaction models and 
depends weakly on the primary energy and mass. 
The muon number density as a function of the position at the ground $\vec{r}$ is then parameterised as $\rho_{\mu}(\vec{r}) = N_{19}~\rho_{\mu,19}(\vec{r};\theta,\phi)$
where $\rho_{\mu,19}(\vec{r};\theta,\phi)$ is a reference distribution conventionally calculated for primary protons at $10^{19}$ eV using the hadronic interaction model QGSJETII-03, and the scale factor $N_{19}$ represents the shower size relative to the normalization
of the reference distribution \cite{n19}.

\begin{figure*}
\includegraphics[width=78mm]{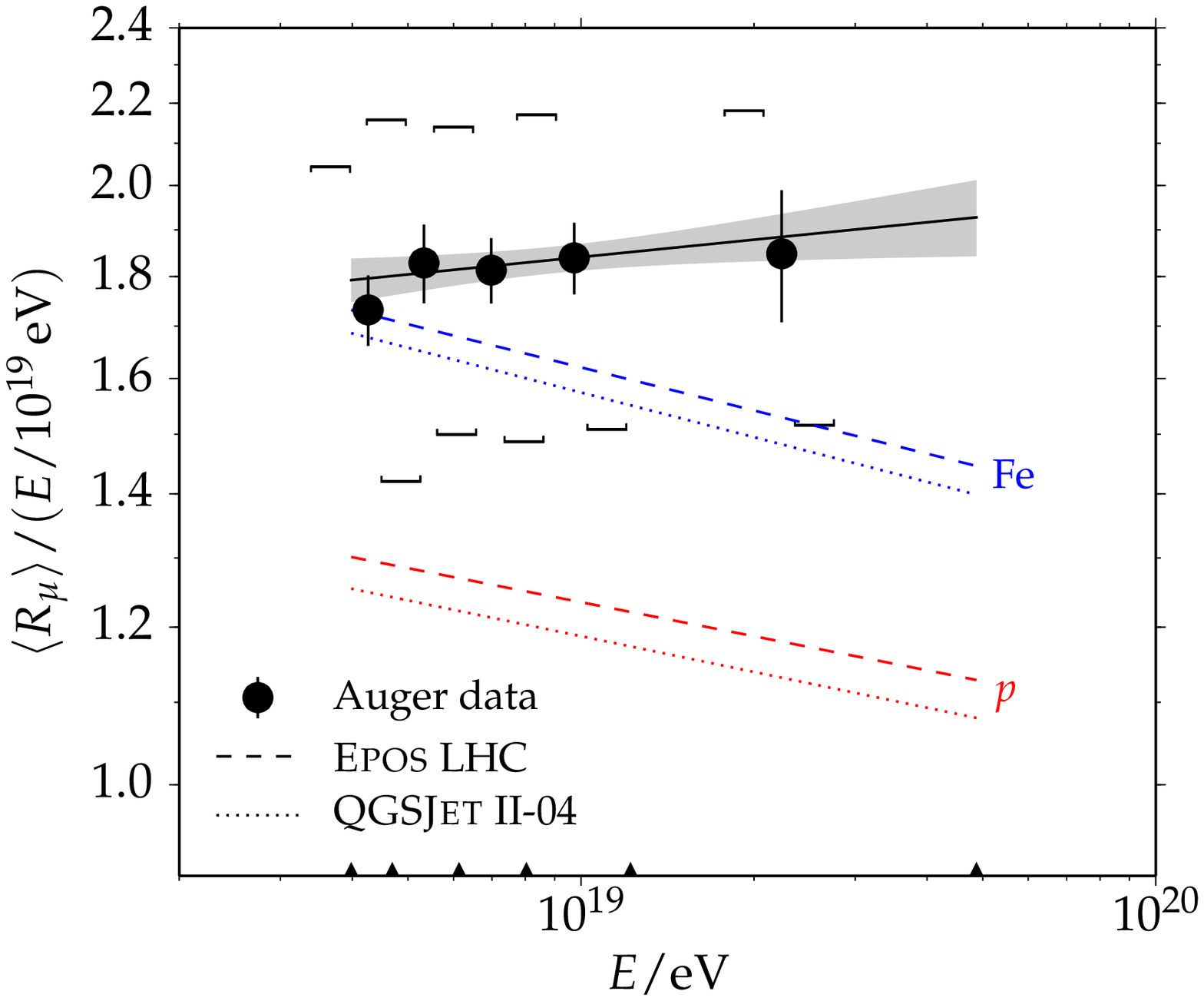}
\includegraphics[width=81mm]{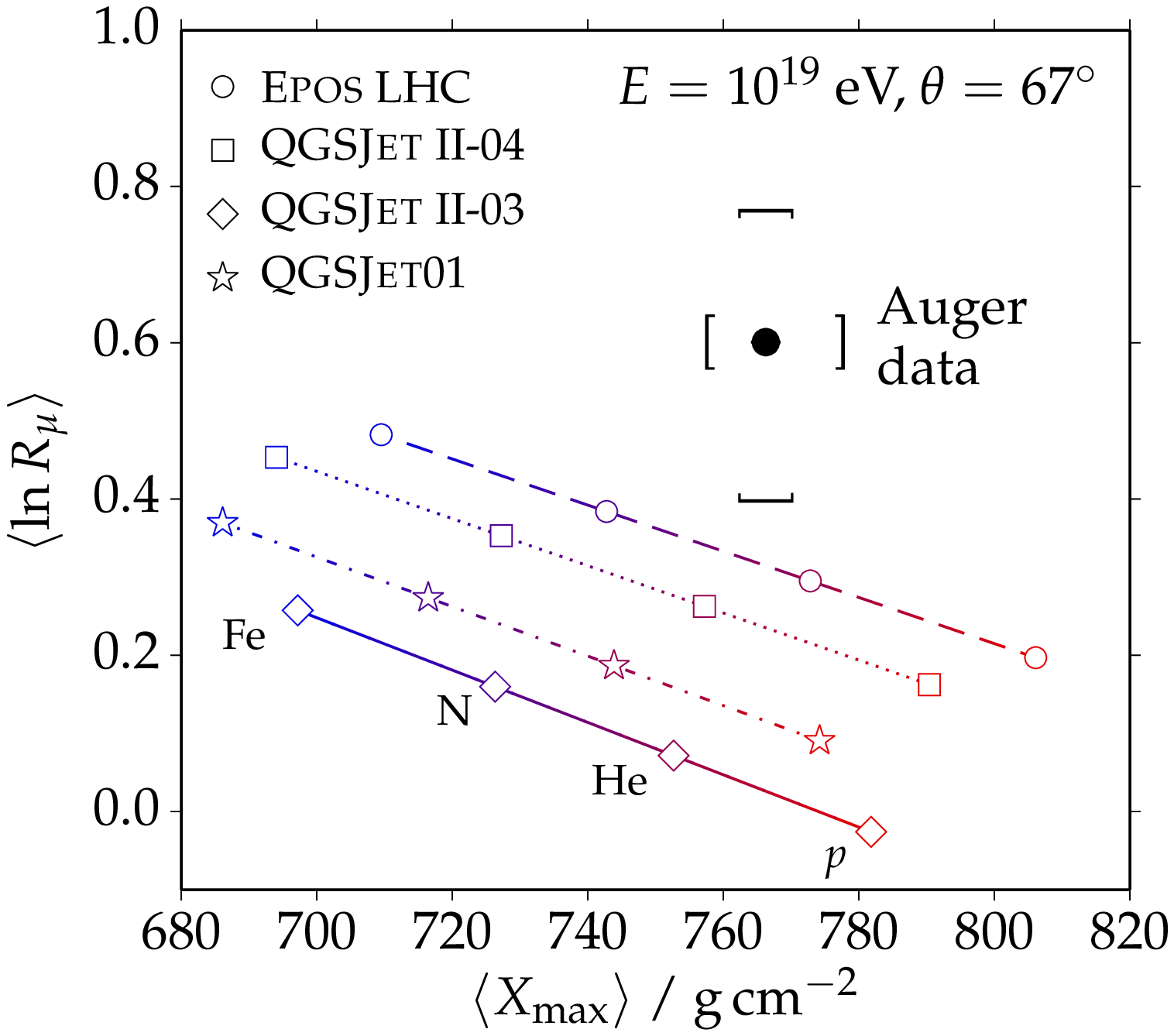}
\caption{Left: Energy-normalized relative muon number $\langle R_\mu \rangle /(E/10^{19}$ eV) for various interaction models and two primaries, p and Fe, as obtained from inclined events \cite{n19}. Note that the Auger data (black points) indicate a heavy composition and exhibit a different slope (black line) than the model predictions (dashed and dotted lines). 
Right: ⟩$\langle \ln  R_\mu \rangle$ vs. $\langle X_\text{max} \rangle$ for various interaction models (lines) and primary masses (markers) and the Auger data (black point with systematic-uncertainty brackets) from events with energies around E = $10^{19}$ eV \cite{n19}.}
\label{fig:models}
\end{figure*}

The average $R_\mu$ over energy bins as a function of energy is shown in Figure \ref{fig:models}-(left), where the energy was reconstructed independently of $R_\mu$ from the FD measurements. The predictions for proton and iron primaries using two hadronic models are also shown for comparison. The systematic uncertainties, depicted with square brackets, are dominated by the uncertainties on the energy scale.

The different value of the measured slope between the data and simulated pure primaries favors a transition from lighter to heavier elements.

Comparing measurements with predictions of simulations, even considering the lowest values of $\langle R_\mu \rangle$ allowed by the systematic uncertainties, the muon content of inclined events would imply a very heavy composition. 

The compatibility of this interpretation with the composition scenario expected from the $X_\text{max}$ measurements is tested in Figure \ref{fig:models}-(right), where 
$\langle \ln R_\mu \rangle$ versus $\langle X_\text{max} \rangle$ at $10^{19}$ eV  is shown. 
One can conclude that, by assuming the $X_\text{max}$ composition, the measured value of $R_\mu$ is significantly larger than that expected from the simulation, regardless of the hadronic model. 
 
The existing models fail to match the measurements and would require substantial increases of muon production rate.

Even though recent hadronic interactions models were tuned to LHC data, there is still room for further improvements.
More precise constraints on the hadronic models will be possible in the future with AugerPrime, the upgrade of the Pierre Auger Observatory. 
\section{The upgrade of the Observatory: AugerPrime}
More than ten years of the Pierre Auger Observatory data changed greatly the view of the community about UHECRs. 
Main results that have been discussed briefly in the previous sections contributed to improve our understanding of UHECRs, but many issues are still debated. 

For example, while the flux suppression above 40 EeV has been observed with more than 20$\sigma$ significance, the origin of this suppression is still unclear. 
The question arises whether the suppression is caused by propagation (GZK effect) or by the exhausted power of the cosmic accelerators.

Also, the measurement of an excess of muons in air showers shows that hadronic interaction models fail to describe sufficiently well all aspects of air showers despite the great improvement in recent years, especially from LHC data. 

A better knowledge of the chemical composition of UHECRs can help to answer these open questions. 

The low duty cycle of the FD does not allow collecting a significant data sample at the highest energies. Several other mass-composition analyses using the SD were performed, but these suffer from larger systematic uncertainties due to the uncertainties in the assessment of the muon content of the shower using the water-Cherenkov detectors. 

To address such challenges, the Pierre Auger Observatory is currently undergoing a major upgrade phase, called AugerPrime.
The main goals are to improve the composition sensitivity of the Pierre Auger Observatory into the flux suppression region and to get an improved measurement of the muonic shower component, with a duty cycle of ∼100 \%.

The key upgrade element is the installation of a thin scintillation detector, dubbed Scintillator Surface Detector (SSD), on top of each of the 1660 water-Cherenkov stations, so that provide complementary information about the electromagnetic and muonic components of the shower. 
Additionally, the duty cycle of the fluorescence telescopes will be extended, allowing a direct determination of the depth of shower maximum with increased statistics at the highest energies.

The electronics of the SD stations is being upgraded to obtain an increased sampling rate and a better timing accuracy, as well as a higher dynamic range, allowing a better reconstruction of the geometry of the showers.
A comprehensive description of the Observatory upgrade can be found in the Preliminary Design Report\cite{dr}.
 
 The first twelve upgraded detector stations of AugerPrime, forming the Engineering Array of the upgrade, were deployed at the Pierre Auger Observatory in 2016.
 They have been in continuous data taking mode since then, producing signals to verify the basic functionality of the detector design.
 The construction of AugerPrime is expected to be finished in 2019. It will be followed by data-taking until 2025 to roughly double the statistics collected so far and add the mass composition information in the region unreachable by the FD.
\section{Conclusions}
The Pierre Auger Observatory continues to provide a wealth of new data of unprecedented quality which have already improved the understanding of UHECRs.

The photon and neutrino limits excluded
the top-down scenarios: UHECRs are accelerated in astrophysical sources.
 
The features of the UHECRs energy spectrum, 
the ankle and the suppression at the highest energies, have been established beyond any doubt.  

A dipole structure in the distribution of arrival directions on Earth has been observed that has an orientation which is indicative of an extragalactic origin of UHECRs. 

The characteristics of the cosmic ray air showers measurements show that current hadronic models still cannot provide a satisfactory description of the muon production in EAS. 

Despite the progresses, a fully consistent interpretation of results is limited by the lack of knowledge at the highest energies. 
To build a consistent picture of the origin of UHECRs, an upgrade, AugerPrime, has been started and will keep the Observatory running at until 2025.
\section*{Acknowledgment} 
The successful installation, commissioning, and operation of the Pierre Auger Observatory would not have been possible without the strong commitment and effort from the technical and administrative staff in Malarg\"ue, and the financial support from a number of funding agencies in the participating countries, listed at \url{https://www.auger.org/index.php/about-us/ funding-agencies}. The corresponding author was supported by CONICET.


\begin{thebibliography}{99}
\bibitem{photon} M. Niechciol for the Pierre Auger Collaboration, PoS ICRC2017 (2018) 517.
\bibitem{neutrino} E. Zas for the Pierre Auger Collaboration, PoS ICRC2017 (2018), 972.
\bibitem{spectrum} F. Fenu for the Pierre Auger Collaboration, PoS ICRC2017 (2018) 486.
\bibitem{mass} J. Bellido for the Pierre Auger Collaboration, PoS ICRC2017 (2018), 506.
\bibitem{anis} Pierre Auger Collaboration, Science, 357(6357):1266–1270, 2017.
 \bibitem{intanis} Pierre Auger Collaboration, Astrophys. J. 853 (2018) no.2, L29.
\bibitem{cross} R.Ulrich for the Pierre Auger Collaboration, PoS ICRC2015, 401.
\bibitem{n19} Pierre Auger Collaboration, Phys. Rev. D 91, 032003 (2015), 1408.1421
\bibitem{dr} Pierre Auger Collaboration, The Pierre Auger Observatory Upgrade AugerPrime: Preliminary Design Report (2016) arXiv:1604.03637
\end{thebibliography}
\end{document}